\documentclass[aps,prb,twocolumn,showpacs,letterpaper]{revtex4}
\usepackage{times}
\usepackage{graphicx}
\usepackage[]{subfigure}
\newcommand{\be}{\begin{equation}}
\newcommand{\ee}{\end{equation}}
\newcommand{\bea}{\begin{eqnarray}}
\newcommand{\eea}{\end{eqnarray}}

\newcommand{\ba}{\begin{array}}
\newcommand{\ea}{\end{array}}
\newcommand{\ie}{{i.e.,}~}
\newcommand{\etal}{{\it et al.}~}

\begin{document}
\title{Absence of jump discontinuity in the magnetization in
  quasi-one-dimensional\\ random-field Ising models} \author{Sanjib
  Sabhapandit} \affiliation{Department of Physics, University of
  Wisconsin, Madison, Wisconsin 53706, USA}

\date[Published in:~]{Phys. Rev. B {\bf 70}, 224401 (2004)}
\begin{abstract}
We consider the zero-temperature random-field Ising model in the
presence of an external field, on ladders and in one dimension with
finite range interactions, for unbounded continuous distributions of
random fields, and show that there is no jump discontinuity in the
magnetizations for any quasi-one dimensional model. We show that the
evolution of the system at an external field can be described by a
stochastic matrix and the magnetization can be obtained using the
eigenvector of the matrix corresponding to the eigenvalue one, which
is continuous and differentiable function of the external field.
\end{abstract}

\pacs{75.10.Nr, 75.60.Ej} 

\maketitle

The nonequilibrium random field Ising model (RFIM) was proposed by
Sethna \etal\cite{SETHNA} as a model for hysteresis and Barkhausen
noise in ferromagnets. Since then, there has been a considerable
theoretical interest in the nonequilibrium response of the
model.~\cite{SETHNA, DHAR, SABHAPANDIT-1, SABHAPANDIT-2, SHUKLA,
  Colaiori, DOBRIN, BLEHER} Sethna \etal showed that in the mean-field
limit, if the strength $\sigma$ of the quenched random field is large,
the average magnetization per site is a continuous function of the
external field, but for small $\sigma$, it shows a discontinuous jump
as the external field is increased. However, in the mean-field limit,
there is no hysteresis above the critical disorder $\sigma_c$, \ie the
magnetization follows the same curve in the increasing and decreasing
field. This shortcoming of the mean-field limit can be overcome by
working on a Bethe lattice, where the the model is solved exactly for
the hysteresis~\cite{DHAR} and avalanche size
distribution.~\cite{SABHAPANDIT-1} Interestingly, the behavior of
hysteresis loops on a Bethe lattice depends nontrivially on the
coordination number $z$, so long as the distributions of the random
fields are unbounded continuous. For $z=3$ the hysteresis loops show
no jump discontinuity of magnetization even in the limit of small
disorder, but for higher $z$ they do. This $z$ dependence of the
hysteresis persists even for euclidean lattices.~\cite{SABHAPANDIT-2}
The natural question to ask is to {\it whether the existence jump
  discontinuity in magnetizations at low disorder depend only on the
  coordination number or also on the dimensionality of the lattice.}
In this paper we answer this question by showing the nonexistence of
jump discontinuity in magnetizations for any quasi-one-dimensional
RFIM, irrespective of the coordination number, for unbounded
continuous distributions of random fields.

\begin{figure}[b]
\centerline{\includegraphics[width=8cm]{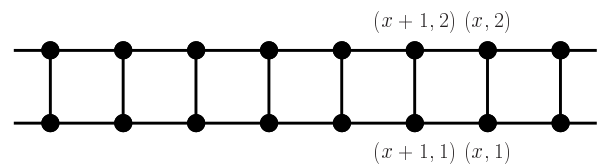}}
\caption{\label{ladder} Two-leg ladder}
\end{figure}

We first consider a two-leg ladder (shown in Fig.~\ref{ladder}) of
length $N$. At each vertex there is a Ising spin $s_i=\pm 1$ which
interacts with nearest neighbors through a ferromagnetic interaction
$J$, and coupled to the on-site quenched random field $h_{i}$ and the
homogeneous external field $h$.  The random fields $\{h_i\}$ are drawn
independently from a unbounded continuous distribution $\phi(h_i)$.
The Hamiltonian of the system is
\be
H=-J \sum_{<i,j>} s_is_j -\sum_{i}h_is_i -h\sum_{i}s_i. 
\label{ladder-H}
\ee 
The system evolves under the zero-temperature Glauber single-spin-flip
dynamics:~\cite{KAWASAKI} a spin flip is allowed only if it lowers the
energy.  We assume that the rate of spin flips is much larger than the
rate at which $h$ is changed, so that all flippable spins may be said
to relax instantly, and any spin $s_i$ always remains parallel to the
net local field $\ell_{i}$ at the vertex
\be 
 s_i= \mbox{sign}(\ell_{i}) = \mbox{sign}( J \sum_{j=1}^{z} s_{j} + h_{i}
+ h) .
\label{LOCAL_FIELD}
\ee 

We start with $h=-\infty$, when all the spins are down, and slowly
increase the field. We are interested in the average magnetization of
the system as a function of external field, away from the two ends of
the ladder in the thermodynamic limit of $N\rightarrow \infty$.

Under the single spin-flip dynamics, for ferromagnetic coupling ($J >
0$), the system exhibits return point memory: if we start with with
the state where all spins are down (at $h=-\infty$), then state of the
system at an increased field $h(T)$ at a later time $T$, does not
depends on the detailed time dependent of the external field $h(t)$,
and is same for all histories as long as the condition $ h(t) \leq
h(T)$ for all earlier times is obeyed.~\cite{SETHNA} Therefore, to
find the magnetization at $h$, we start with $h=-\infty$, so that all
spins in the system are down and then increase the external field to
$h$ in a single step. At that field several spins would become
unstable. But for $J>0$, if we start with any stable configuration,
and then increase the external field and allow the system to relax,
the final stable configuration reached is independent of the order in
which the unstable spins are flipped, as flipping of a spin can only
increase the local field at its neighboring vertices and also in the
relaxation process no spin flips more than once. Because of this
abelian property of the spin flip, when the external field is
increased from $-\infty$ to $h$ in a single step, we may choose to
relax the spins from the two ends of the ladder as follows: We first
relax the spins at the boundary ends (level 1). Then we relax spins
next to the boundaries (at level 2) and so on with increasing levels
as we move towards the center of the ladder. In the relaxation
process, if a spin at level $x$ flips up, we check all the spins at
lower levels for possible upward flips, before flipping the second
spin at level $x$. Note that the state of the spins at a level $r$
from the right end is independent of the state at level $l$ from the
left end, so long as all the spins in between levels are kept down.
Therefore we can relax the spins from the two boundary ends on the
left and right halves of the ladder independently (which are
identical).

We choose our coordinates on each halves of the ladder such that the
level $x$ consists of vertices $(x,1)$ and $(x,2)$.  Let
$P^{x}_h(s_1,s_2)$ be the conditional probability that, in the earlier
spin relaxation scheme, the spins at level $x$ reach a final state
$\{s_1,s_2\}$, \ie the spin at the vertex $(x,1)\rightarrow s_1$ and
the spin at the vertex $(x,2)\rightarrow s_2$; given that the spins at
level $x+1$ are kept down and all the spins at the lower levels are
relaxed and the external field is $h$.  Corresponding to four possible
final states we get four probabilities $P^x_h(1,1)$, $P^x_h(1,-1)$,
$P^x_h(-1,1)$ and $P^x_h(-1,-1)$, which depend on the states at level
$x-1$ and the random fields at level $x$.  Note that
$P^x_h(1,-1)=P^x_h(-1,1)$, due to the symmetry of the lattice.

To illustrate how to obtain the recursions for the probabilities
$P^x_h(s_1,s_2)$'s, we consider an example where the final $\{1,1\}$
state at level $x$ is achieved from its initial state $\{-1,-1\}$,
when state at level $x-1$ is \{1,1\}.  We denote $p_m(h)$ with $0 \leq
m \leq 3$ as the conditional probability that the local field at any
vertex $i$ will be large enough so that it will flip up, if $m$ of its
neighbors are up, when the uniform external field is $h$.  Clearly,
for a given distribution of random fields $\phi(h_i)$:
\be
p_{m}(h)=\int_{(3-2m)J-h}^{\infty} \phi(h_{i}) \, \,dh_{i},~~~~0\le m\le3.
\label{p_m}
\ee 
Since the spin at vertex $(x,1)$ has one up neighbor, it will flip up
with probability $p_1(h)$. Now the spin at vertex $(x,2)$ has two up
neighbors, so it flips up with probability $p_2(h)$.  But the local
field at vertex $(x,1)$ may not be large enough for the spin to flip
up when it has only one up neighbor. In this case, the spin at vertex
$(x,2)$ flips up first, with probability $p_1(h)$, and then the spin
at vertex $(x,1)$ flips up, when is has two up neighbors, with
probability $[p_2(h)-p_1(h)]$.  Therefore, the total probability of
the spins at level $x$ flipping up via this process is
$P^{(x-1)}_h(1,1) \{p_1(h) p_2(h) + [p_2(h)-p_1(h)] p_1(h)\}$.

The case where the state at level $x-1$ is $\{-1,-1\}$, the flipping
of the spin at vertex $(x,1)$ might causes the spin at vertex
$(x-1,1)$ to flip up and as a result the spin at vertex $(x-1,2)$
might flip up, while the spin at vertex $(x,2)$ is still kept down. We
denote the probability of this spin flip process by $Q_h^{(x-1)}$. If
we consider $[P^x_h(1,1), 2P^x_h(-1,1), P^x_h(-1,-1)-Q^x_h,Q^x_h]$ as
a column vector $\mathcal{P}_x(h)$, then the recursion relations for
these probabilities can be represented in matrix form
\be 
\mathcal{P}_x(h) = W_h \mathcal{P}_{x-1}(h),
\label{recursion}
\ee
where $W_h$ is a $4\times4$ matrix whose elements $w_{ij}$'s are
polynomials in $p_m(h)$'s.

For a given probability distribution of random fields $\phi(h_i)$ and
value of external field $h$, we determine $p_m(h)$, and then, using
Eq.~(\ref{recursion}) we can recursively determine the probabilities
represented by the column vector $\mathcal{P}_x(h)$.  Note that the
matrix $W_h$ is column stochastic,~\cite{BERMAN} \ie $w_{ij}\ge 0$ and
$\sum_{i=1}^4 w_{ij}=1$. Therefore, for large $x$, the vector
$\mathcal{P}_x(h)$ tends to a limiting vector $\mathcal{P}^\star(h)$,
which is the eigenvector of the matrix $W_h$, corresponding to the
eigenvalue one (the maximal eigenvalue). The variation of
$\mathcal{P}^\star(h)$ with respect to the external field $h$
is~\cite{GOLUB}
\be
\frac{d\mathcal{P}^\star(h)}{dh}=
A_h^\#\frac{dW_h}{dh}\mathcal{P}^\star(h),
\ee
where $A_h^\#$ is the group inverse of the matrix $A_h=I-W_h$, letting
$I$ denote identity matrix.  The entries of $A_h^\#$ are continuous
functions of $h$ for continuous distributions of random fields. Since
according to Eq.~\ref{p_m}, the elements of $W_h$ are continuous and
differentiable (with continuous first derivative),
$\mathcal{P}^\star(h)$ is a continuous and differentiable (with
continuous first derivative) function of $h$.

To calculate the magnetization at the center vertex $(O,1)$, in the
limit $N\rightarrow \infty$, we keep the spins at level $O$ down and
relax the full system. The states of the spins at the adjacent levels
on both sides are given independently by the probability vector
$\mathcal{P}^\star(h)$. Now we relax the spins at at level $O$. Let
$P^O_h(-1,-1)$ be the probability that in this relaxation process the
spins at level $O$ do not flip up and $P^O_h(-1,1)$ be the probability
that only the spin at vertex $(O,2)$ flips up.  The magnetization at
the vertex $(O,1)$ is
\be M_O(h)=1-2 \left[P^O_h(-1,-1) + P^O_h(-1,1)\right].
\label{magnetization-ladder}
\ee
In the limit $N\rightarrow \infty$, all the vertices deep inside the
ladder are equivalent. Therefore, $M_O(h)$ gives the average
magnetization on the ladder far away from the boundary. Since for
continuous distributions of random fields $\mathcal{P}^\star(h)$ is
continuous and differentiable, the average magnetization is also
continuous and differentiable function of $h$.

\begin{figure}
\centerline{\includegraphics[width=8cm]{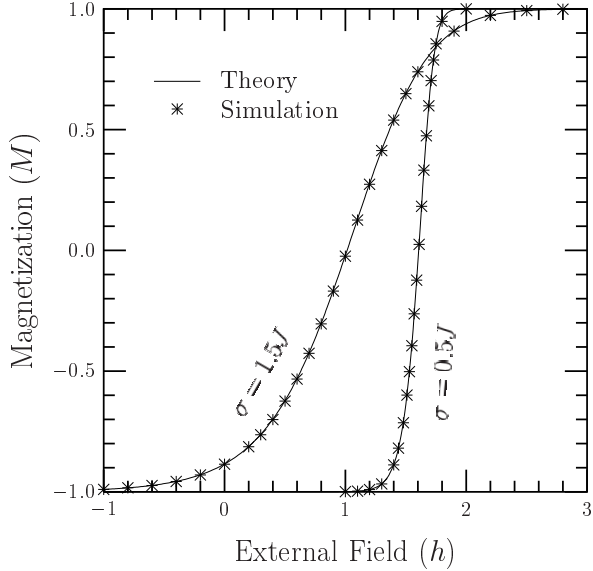}}
\caption{\label{mag-ladder} Magnetization curves for a two-ladder in the
  increasing field. $N=2^{23}$.}
\end{figure}

We have compared the theoretical calculation with numerical simulation
results. The points in Fig.~\ref{mag-ladder} show the results of
simulations for a single realization of quenched Gaussian random
fields with mean $0$ and standard deviation $\sigma=0.5J$ and
$\sigma=1.5J$, respectively. We use periodic boundary condition along
the length of a ladder with $N=2^{23}$. Different runs using different
realizations of quenched field give results which are
indistinguishable at the scale of the graph. The solid curves the
Fig.~\ref{mag-ladder} are obtained using theoretical calculation.

The analysis of a two-leg ladder can be extended to a case of an
$n$-leg ladder (with finite $n$), where the relaxation scheme is such
that, all the $n$ spins at the same level $x$ are relaxed before
relaxing the spins at level $x+1$.  The relaxation process can be
represented by the probabilities $P_h^x(s_1,s_2,\ldots, s_n)$ of the
states of spins at level $x$ and the probabilities $Q_i^x(h)$'s for
intermediate relaxations. Now these probabilities at level $x$ can be
obtained from the probabilities at level $x-1$, and using proper
linear combinations these can be expressed by a matrix recursion
relation like Eq.~(\ref{recursion}). It is simpler to consider the
$n$-leg ladder to wrapped around a cylinder (of course the resultant
magnetization is quantitatively different from the case of an $n$-leg
ladder, but the qualitative behavior remains unchanged), since then
many of the probabilities become equal, due to the rotational symmetry
of the cylinder. As an example, consider the case $n=3$. The state at
a level $x$, when the spins at level $x+1$ are kept down, can be
represented by only using $P_h^x(1,1,1)$, $P_h^x(1,1,-1)$,
$P_h^x(1,-1,-1)$, and $P_h^x(-1,-1,-1)$. The $Q_i(h)$'s needed to
describe the relaxation at level $x-1$, when relaxing the spins at
level $x$ are obtained as follows:

(a) Suppose the state at level $x-1$ is $\{1,-1,-1\}$ and the spin at
vertex $(x,2)$ flips up and the spin at vertex $(x,3)$ is kept
down. We define $Q_1^{(x-1)}(h)$ to be the probability that the spin
at vertex $(x-1,2)$ flips up and subsequently the spin at vertex
$(x-1,3)$ flips up as a result. (b) Suppose the state at level $x-1$
is $\{-1,-1,-1\}$ and the spin at vertex $(x,1)$ flips up while the
spins at vertices $(x,2)$ and $(x,3)$ are kept down. Now we define:
(i) $Q_2^{(x-1)}(h)$, the probability that the spin at vertex
$(x-1,1)$ flips up and as a result the spins at vertices $(x-1,2)$ and
$(x-1,3)$ flip up. (ii) $Q_3^{(x-1)}(h)$, the probability that the
spin at vertex $(x-1,1)$ flips up and as a result the spin at vertices
$(x-1,2)$ flips up but the spin at vertex $(x-1,3)$ remains
down. (iii) $Q_4^{(x-1)}(h)$, the probability that the spin at vertex
$(x-1,1)$ flips up but the spins at vertices $(x-1,2)$ and $(x-1,3)$
remain down, and (iv) $Q_5^{(x-1)}$, the probability that the spins at
vertices $(x-1,2)$ and $(x-1,3)$ flip up after we flip the spin at
vertex $(x,2)$ and the spin at vertex $(x,3)$ is still kept down.

Now the elements of the probability vector $\mathcal{P}_x(h)$ are
$P_h^x(1,1,1)$, $3P_h^x(1,1,-1)$, $3[P_h^x(1,-1,-1)-Q_1^x(h)]$,
$3Q_1^x(h)$, $P_h^x(-1,-1,-1)-Q_2^x(h)-Q_3^x(h)-Q_4^x(h)$, $Q_2^x(h)$,
$Q_3^x(h)$, $Q_4^x(h)-Q_5^x(h)$ and $Q_5^x(h)$. The recursion relation
for $\mathcal{P}_x(h)$ is represented by Eq.~\ref{recursion} with a
$9\times9$ stochastic matrix $W_h$ whose elements can be expressed in
terms of $p_m(h)$'s. Finally, the magnetization can be obtained using
the limiting vector $\mathcal{P}^\star(h)$ and $p_m(h)$'s.

We next consider the RFIM in one dimension (1D), with ``finite range
interactions'' --- each spin interacts with all the spins up to a
distance $n$, through ferromagnetic interaction $J$. We take the
length of the chain $N$ (later we will take the limit $N\rightarrow
\infty$) to be a multiple of $n$. Now we can group $n$ spins together
in a single level and relax the spins from the two boundaries such
that we relax all the spins at the lower levels before relaxing the
spins at an higher level.

\begin{figure}
\centerline{\subfigure[]{\includegraphics[width=8cm]{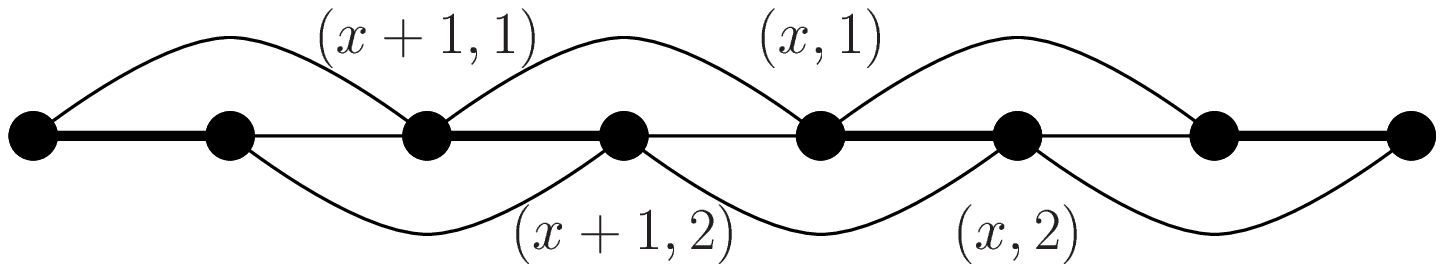}}}
\centerline{\subfigure[]{\includegraphics[width=8cm]{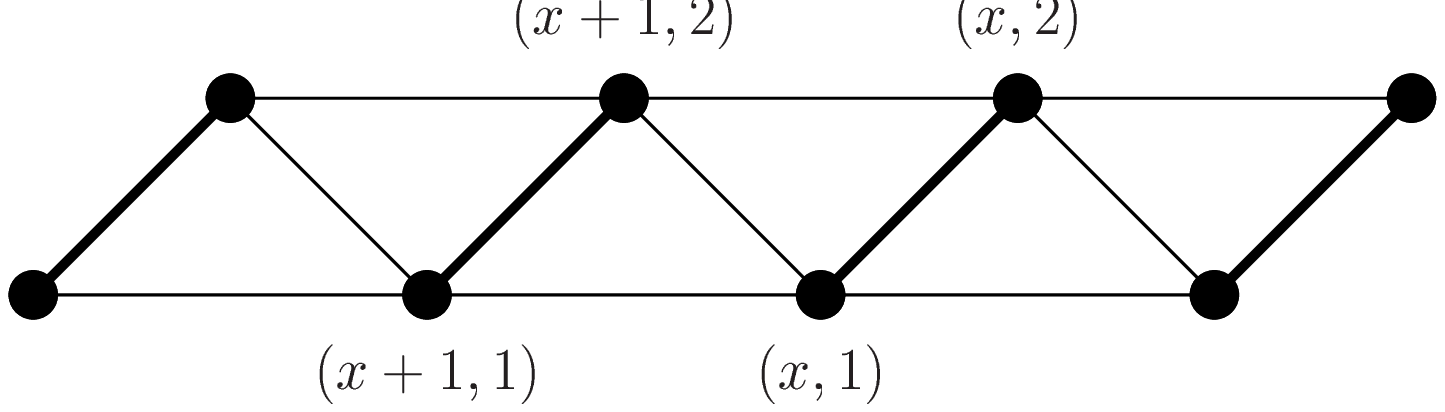}}}
\caption{\label{next-neighbor}(a) Linear chain showing nearest and
  next neighbor interactions and (b) graph on which nearest neighbor
  interaction is equivalent to the linear chain with nearest and next
  neighbor interactions.}
\end{figure}

As a concrete example, we calculate the magnetization for $n=2$, \ie
each spin interacts with the nearest and next nearest neighbors. The
interactions are shown in Fig.~\ref{next-neighbor}(a). This model is
also equivalent to the RFIM with nearest neighbor interaction on the
graph shown in Fig.~\ref{next-neighbor}(b). Let $P^{x}_h(s_1,s_2)$ be
the conditional probability that the spins at level $x$ reach a final
state $\{s_1,s_2\}$, \ie the spin at the vertex $(x,1)\rightarrow s_1$
and the spin at the vertex $(x,2)\rightarrow s_2$; given that the
spins at level $x+1$ are kept down and all the spins at the lower
levels are relaxed. We need to define four more probabilities for the
further relaxation at level $x-1$, as a result of a spin flip at level
$x$.

Consider the following cases of relaxing the spins at level $x$ from
its original state $\{-1,-1\}$: (a) Suppose the state at level $x-1$
is $\{-1,1\}$ and the spin at vertex $(x,2)$ flips up, given that
spins at vertex $(x,1)$ and level $x+1$ are kept down. It may cause
the spin at vertex $(x-1,1)$ to flip up and we denote the probability
of it to flip up by $Q_1^{x-1}(h)$. The spin at vertex $(x-1,1)$
remains down with probability $P_h^{x-1}(-1,1)-Q_1^{x-1}(h)$.  (b)
Suppose the state at level $x-1$ is $\{-1,-1\}$ and the spin at vertex
$(x,2)$ flips up, given that spins at vertex $(x,1)$ and level $x+1$
are kept down. This may finally cause the spin at vertex $(x-1,1)$ to
flip up. The spin at vertex $(x-1,2)$ may or may not flip up during
the relaxation. We denote the probability of the spin at vertex
$(x-1,1)$ flipping up by $Q_2^{x-1}(h)$. It remains down with
probability $P_h^{x-1}(-1,-1)-Q_2^{x-1}(h)$.  (c) Suppose the state at
level $x-1$ is $\{-1,-1\}$ and the spin at vertex $(x,1)$ flips up,
given that spins at vertex $(x,2)$ and level $x+1$ are kept down. Now
we relax the spins at level $x-1$.  The spin at vertex $(x-1,2)$ can
not flip up in this further relaxation unless the spin at vertex
$(x-1,1)$ flips up first. Let $Q_3^{x-1}(h)$ be the probability that
the spin at vertex $(x-1,1)$ flips up, but it can not cause the spin
at vertex $(x-1,2)$ to flip up. We denote the probability of both the
spins at level $x-1$ flipping up by $Q_4^{x-1}(h)$. Therefore, the
spins at level $x-1$ remain down with probability
$P_h^{x-1}(-1,-1)-Q_3^{x-1}(h)-Q_4^{x-1}(h)$.

\begin{figure}
\centerline{\includegraphics[width=8cm]{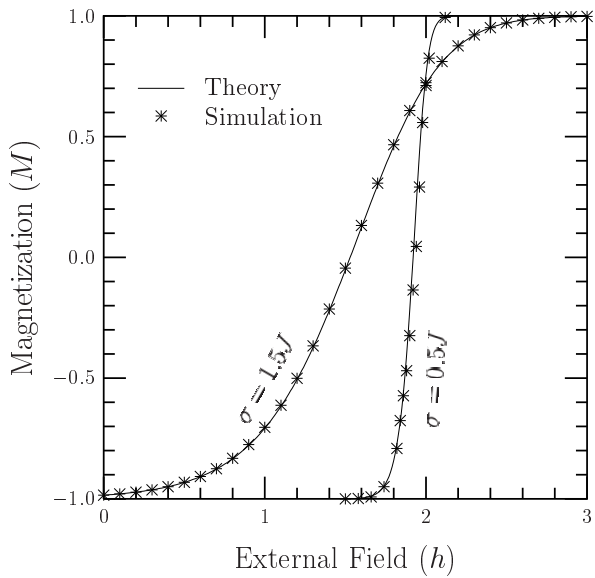}}
\caption{\label{mag-nn} Magnetization curve in the increasing field
  for linear chain with nearest and next neighbor
  interactions. $N=2^{23}$.}
\end{figure}

We consider the probabilities [$P_h^x(1,1)$, $P_h^x(1,-1)$,
$P_h^x(-1,1)-Q_1^x(h)$, $P_h^x(-1,-1)-Q_2^x(h)$, $Q_1^x(h)$,
$Q_2^x(h)-Q_3^x(h)-Q_4^x(h)$, $Q_3^x(h)$, $Q_4^x(h)$] as a column
vector $\mathcal{P}_x(h)$, so that the recursion relations of the
probabilities can be represented in matrix form as in
Eq.~\ref{recursion}, with a $8\times8$ stochastic matrix $W_h$. We
find the eigenvector $\mathcal{P}^\star(h)$ for the matrix $W_h$ and
using $\mathcal{P}^\star(h)$ and $p_m(h)$'s calculate the average
magnetization far away from the boundary.  Figure~\ref{mag-nn} shows
the comparison between theoretical and simulation results for Gaussian
quenched random fields with $\sigma=0.5J$ and $\sigma=1.5J$. The
simulation results are obtained for single realizations, on a linear
chain of length $N=2^{23}$ with nearest and next neighbor
interactions, with periodic boundary condition.

In summary, we have demonstrated that there is no jump discontinuity
in the magnetizations for any quasi-one dimensional RFIM for unbounded
continuous distributions of random fields.  We showed that for the
RFIM on an $n$-leg ladder and in 1D with interactions extended to $n$
closest neighbors, for finite $n$ the relaxation of spins at an
external field $h$ can be described by a stochastic matrix $W_h$; and
the average magnetization can be obtained using the eigenvector of
$W_h$ corresponding to the eigenvalue one, which is a continuous and
differentiable function of $h$, for unbounded continuous distributions
of random fields.  We explicitly calculated the magnetizations for two
simpler cases: (a) the two-leg ladder and (b) in 1D with nearest and
next nearest neighbors interactions; and confirmed our results using
numerical simulation for Gaussian distribution of random fields.  The
question of how to take the $n\rightarrow \infty$ limit, where the
magnetizations show jump discontinuity below a critical disorder as
the external field is varied, remains open.

The author would like to thank D. Dhar and S.~N. Coppersmith for their
useful comments on the manuscript. This work was supported by National
Science Foundation grant No.~DMR-0209630.


\begin{thebibliography}{99}
\bibitem{SETHNA} J. P. Sethna, K. A. Dahmen, S. Kartha, J. A.
  Krumhansl, B. W. Roberts, and J. D. Shore, Phys. Rev. Lett. {\bf 70},
  3347 (1993).
  
\bibitem{DHAR} D. Dhar, P. Shukla, and J. P. Sethna, J Phys A {\bf
    30}, 5259 (1997).
  
\bibitem{SABHAPANDIT-1} S. Sabhapandit, P. Shukla, and D. Dhar, J.
  Stat. Phys. {\bf 98}, 103 (2000).
  
\bibitem{SABHAPANDIT-2} S. Sabhapandit, D. Dhar, and P. Shukla, Phys.
  Rev. Lett. {\bf 88}, 197202 (2002).
  
\bibitem{SHUKLA} P. Shukla, Phys. Rev. E {\bf 62}, 4725 (2000); {\bf
    63}, 027102 (2001).

\bibitem{Colaiori} F. Colaiori, A. Gabrielli, and S. Zapperi,
  Phys. Rev. B {\bf 65}, 224404 (2002).
  
\bibitem{DOBRIN} R. Dobrin, J. H. Meinke, and P. M. Duxbury, J Phys A
  {\bf 35}, L247 (2002).
  
\bibitem{BLEHER} P. M. Bleher, J. Ruiz, and V. A. Zagrebnov, J. Stat.
  Phys. {\bf 93}, 33 (1998), and references therein.
  
\bibitem{KAWASAKI} See, for example: K. Kawasaki, {\em Phase
    Transitions and Critical Phenomena}, edited by C. Domb and M. S.
  Green (Academic Press, London, 1972), vol 2.

\bibitem{BERMAN} A. Berman and R. J. Plemmons, {\em Nonnegative
  Matrices in the Mathematical Sciences} (Academic Press, New York,
  1979).
  
\bibitem{GOLUB} G. H. Golub and C. D. Meyer, SIAM J. Algebraic
  Discrete Methods {\bf 7}, 273 (1986); C. D. Meyer, SIAM J. Matrix
  Anal. Appl. {\bf 15}, 715 (1994).

\end{thebibliography}
\end{document}